\documentclass[12pt,preprint]{aastex}
\usepackage{colordvi}
\usepackage{color}

\newenvironment{tight_enumerate}{
\begin{enumerate}
  \setlength{\itemsep}{0pt}
  \setlength{\parskip}{0pt}
}{\end{enumerate}}

\begin{document}

\title{Collective Study of Polar Crown Filaments in the Past Four Solar Cycles}

\author{Yan Xu\altaffilmark{1,2},
        Werner P{\"o}tzi\altaffilmark{3},
        Hewei Zhang\altaffilmark{4},
        Nengyi Huang\altaffilmark{1},
        Ju Jing\altaffilmark{1,2},
        and
        Haimin Wang\altaffilmark{1,2}}

\affil{1.\ Space Weather Research Lab, Center for
Solar-Terrestrial Research, \\ New Jersey Institute of Technology \\
323 Martin Luther King Blvd, Newark, NJ 07102-1982, USA}

\affil{2.\ Big Bear Solar Observatory, \\ New Jersey Institute of Technology \\
40386 North Shore Lane, Big Bear City, CA 92314-9672, USA}

\affil{3.\ Kanzelh{\"o}he Observatory for Solar and Environmental Research, \\ University of Graz, \\ Kanzelh{\"o}he 19, 9521 Treffen am Ossiacher See, Austria}

\affil{4.\ Department of Mathematical Sciences, \\New Jersey Institute of Technology \\
323 Martin Luther King Blvd, Newark, NJ 07102-1982}

\date{\today}

\clearpage

\begin{abstract}

    Polar Crown Filaments (PCFs) form above the magnetic polarity inversion line, which separates the unipolar polar fields and the nearest dispersed fields from trailing part of active regions with opposite polarity. The statistical properties of PCFs are  correlated with the solar cycle. Therefore, study of PCFs plays an important role in understanding the variation of solar cycle, especially the prolonged cycle 23 and the current `abnormal' solar cycle 24. In this study, we investigate PCFs using full disk H$\alpha$ data from 1973 to early 2018, recorded by Kanzelh\"{o}he Solar Observatory (KSO) and Big Bear Solar Observatory (BBSO), in digital form from 1997 to 2018 and in 35 mm film (digitized) from 1973 to 1996. PCFs are identified manually because their segmented shape and close-to-limb location were not handled well by automatical detections in several previous studies. Our results show that the PCFs start to move poleward at the beginning of each solar cycle. When the PCFs approach to the maximum latitude, the polar field strength reduces to zero followed by a reversal. The migration rates are about 0.4 to 0.7 degree per Carrington rotation, with clear N-S asymmetric pattern. In cycles 21 and 23, the PCFs in the northern hemisphere migrate faster than those in the southern hemisphere. However, in the `abnormal' cycle 24, the southern PCFs migrate faster, which is consistent with other observations of magnetic fields and radio emission. In addition, there are more days in cycle 23 and 24 without PCFs than in the previous cycles.

\end{abstract}


\section{Introduction}

Polar crown filaments (PCFs) are formed above the polarity inversion line (PIL) on the
boundary of polar crown cavity, between the dispersed magnetic field and the predominate
polar field of the previous cycle \citep{Leroy1983, Hirayama1985, Tang1987, Panesar2014}.
\citet{Hansen1975} characterized filaments into three categories, according to their
latitudes, namely low-latitude active region filaments, mid-latitude diagonal filaments
and PCFs. The former two types of filaments migrate together with major sunspot groups
toward solar equator, but the PCFs were found `rush to the pole' \citep{Lockyer1931,
Hyder1965}. According to the solar dynamo models, the cyclic change is driven by the emergence and decay of concentrated magnetic flux \citep{Parker1955}. While the leading sunspots migrate toward equatorial latitudes, the dispersed magnetic fields of following polarity move toward and cancel with the polar fields \citep{Leighton1964, Hathaway1996, Wang1991, Dikpati2001, Simoniello2016}. As a consequence, the polar crown cavity shrinks and PCFs drift poleward together with their PILs. Related to the magnetic cancellation, CME associated PCF eruptions may play an important role in the polar reversal. \citet{Gopalswamy2003a} argued that filaments are in the form of closed fields and their eruptions would remove these closed fields, and consequently open fields can be acquired in the pole for the polar field reversal.

Periodic migration of filaments, including both equatorward and poleward, has been reported and studied qualitatively by many papers, such as \citet{Ricco1892,Ananthakrishnan1952,Fujimori1984,Lorenc2003,Cliver2014,Gopalswamy2018}. \citet{Hansen1975} studied the latitudinal distribution of all filaments during the solar cycle $\sharp$20. Stand alone PCF ($40^{o} \sim 80^{o}$ latitude) migration between 1956 to 1963 (solar cycle $\sharp$19) was investigated by \citet{Hyder1965}. The authors showed distinct horizontal orientation of PCFs and found N-S asymmetry of filament distributions, which is similar to that found by \citet{Waldmeier1971} in the previous cycle $\sharp$19. In a following paper, \citet{Hansen1975a} proposed magnetic configurations for the ascending and descending phases of solar cycles and discussed the disappearance of PCFs at solar maximum. \citet{Tlatov2016} analyzed the tilt angle of filament and found PCFs have the smallest tilt angle, in agreement with the near horizontal PILs described by \citet{Leroy1983}. \citet{Hao2015} presented an automatic method in detecting filaments using the full-disk H$\alpha$ data obtained by Big Bear Solar Observatory (BBSO), from 1988 to 2013. They found that most filaments ($> 80\%$) have tilt angles less than $60^{o}$ and the opposite N-S asymmetries dominated solar cycles $\sharp$22 and $\sharp$23.
\citet{Chatterjee2017} studied H$\alpha$ maps in a longer period from 1914 to 2007, taken by Kodaikanal Solar Observatory. They found an offset between the periodic variation of PCFs and polar magnetic fields, which is consistent with the result in \citet{Gopalswamy2018}. As we will discuss below, we found that the automatic detection codes may cause significant uncertainties in identifying PCFs. As a result, the properties of PCFs were not described clearly and accurately in previous studies.

In this paper, we present the study of PCFs, observed by Kanzelh{\" o}he Solar Observatory
(KSO) and BBSO, from 1973 to early 2018. PCFs are identified visually, as the current automatic detection methods usually treat segmented long filaments as multiple individual ones and may miss the filaments near the limb due to the reduced image contrast. We study the poleward migration rate and the N-S asymmetry of PCFs during each of the 4 solar cycles. In addition, we compare the distribution and migration of PCFs with magnetic fields to investigate the correlation between PCF migration and the peak of sunspot numbers.

\section{Data Mining and Results}

\begin{figure}
\centering
\includegraphics[scale=0.88]{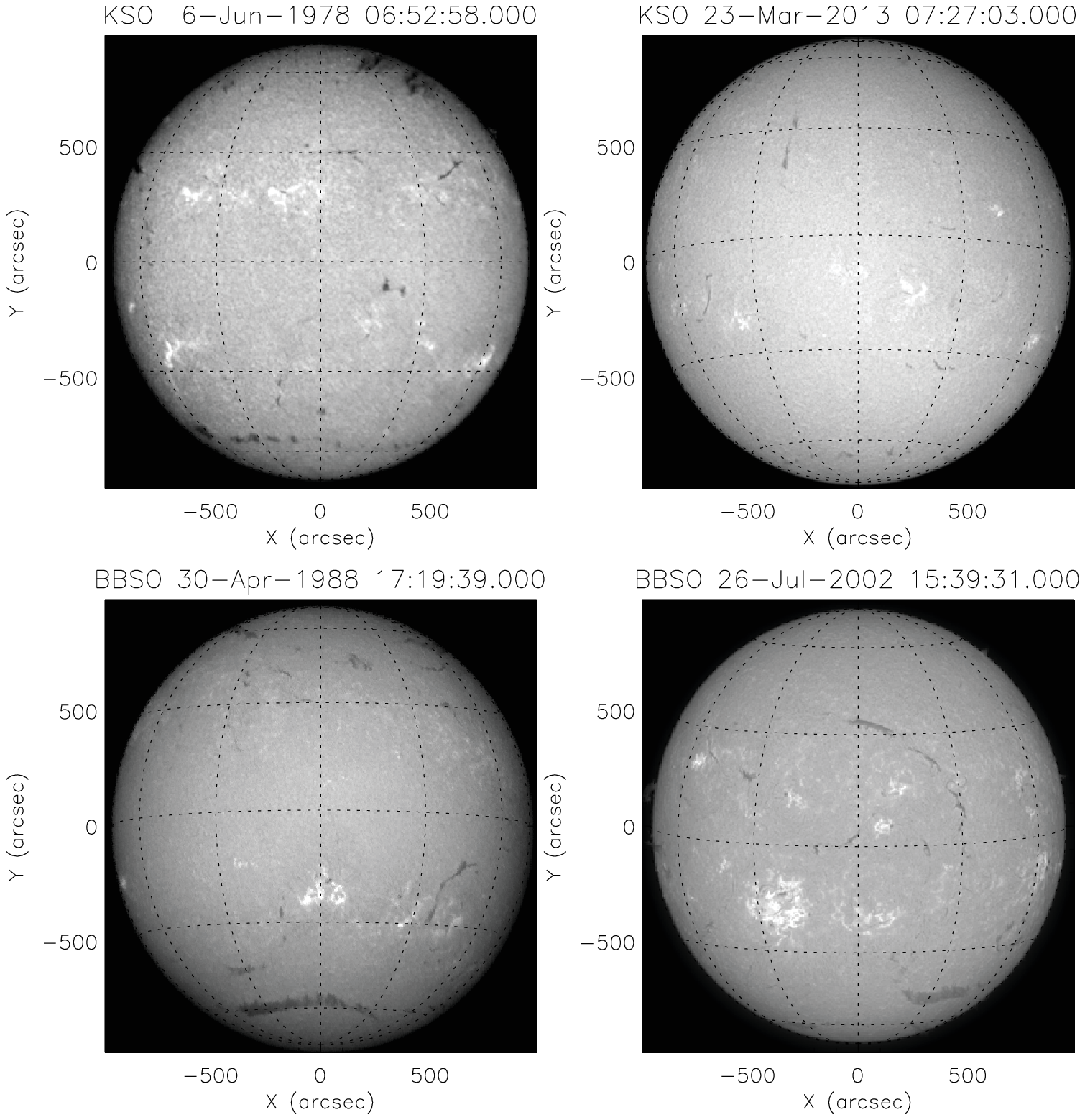}
\caption{Film(left) and digital (right) taken by KSO (top) and BBSO (bottom), respectively.
\label{dataexmp}}

\end{figure}

KSO and BBSO have long history in H$\alpha$ observations, which are favorable data sets for filament detection because of the suitable formation temperature of about 10000 K \citep{Wang1998}. The data set analyzed here starts from 1973 May to the most current images obtained in 2018 March. The original archive consists two types of data, recorded by CCDs (refers to digital data in the rest of this paper) and recorded by 35-mm films (refers to film data), in which the latter has been digitized for scientific purposes. Figure~\ref{dataexmp} shows examples of film and digital data taken by KSO and BBSO, respectively. The spatial and temporal resolutions are typically 1\arcsec ($\sim$ 725 km) and 1 min, respectively. PCFs usually appear on the disk for several weeks or months, we used the daily images instead of the high cadence 1-min data. The daily images are selected
manually among the images of highest quality of the observation day. The digital data is in a common format of FITS, which can be directly registered to the heliographic coordinates. For the film data, geographic parameters, such as disk center location and diameter, are determined by finding the solar limb with the highest intensity gradient against the background. The digital data has been corrected for dark and flat fields, which are not available for the film data. In fact, filaments are usually very obvious on H$\alpha$ images, therefore the nonlinearity of the film data does not affect the identification of PCFs.

The magnetograms are obtained by the 512-Channel Magnetograph and Spectromagnetograph at
the National Solar Observatory at Kitt Peak (NSO/KP), the Michelson Doppler Imager (MDI)
onboard Solar and Helopspheric Observatory (SOHO) and the Helioseismic and Magnetic Imager
(HMI) onboard the Solar Dynamics Observatory (SDO). NSO/KP provides synoptic chats starting
from 1975 February (Carrington rotations No. 1625) in FITS format. The averaged magnetic
field strength, in the direction of line-of-sight (LOS), near the north and south poles
can be measured from these synoptic maps. For MDI and HMI data, the radial polar fields
are provided, in a cadence as high as 720 s \citep{Sun2015}. To avoid contamination of
the polar field by flux concentrations in high latitudes, we simply used the radial polar
fields above $50^{o}$, which can be downloaded directly from HMI's data center (http://jsoc.stanford.edu/ajax/lookdata.html).

In recent years, several methods have been developed for automatic detection of filaments
\citep{Qu2005, Yuan2011, Potzi2015, Hao2015}. On the other hand, however, the visual
detection still provides better accuracy, especially for the segmented PCFs. Furthermore,
visual detection does not require uniform calibration for different data sets. Therefore,
in this study we chose to identify PCFs manually, according to several criteria:

\begin{tight_enumerate}

\item[1.] Filaments near ARs or obvious between two plage regions are excluded. PCFs form  above the PILs that separate polar cavity and diffused fields. In H$\alpha$, the diffused fields appear to be weak plage areas. Therefore, PCFs can not be formed between two belts of diffused fields/plage areas.
\item[2.] Filaments anchored on ARs are excluded.
\item[3.] Filaments with large negative tilt angles ($>$ 25 degrees) or with positive tilt angles are excluded. This is because that the tilt angle of PCFs are found to be near zero or negative \citep{Tlatov2016}.
\item[4.] For weak and segmented filaments, we monitor them for several days. A positive identification will be made if such a filament appears at the limb obviously or its trailing part becomes significant.
\item[5.] Filaments located above $40^{o}$ are counted \citep{Babcock1955}. This threshold is a compromise of other thresholds specified in the literature \citep{Brajsa1991, Hao2015}.

\end{tight_enumerate}
Eventually, in total about 700 PCFs are identified. In order to minimize the
subjective uncertainty introduced by visual detection, multiple operators are involved for mutual cross-check. Our method is time consuming, but currently provides more reliable  detection of PCFs, especially for those are segmented and close to the solar limb. It minimizes the contamination of other types of filaments and therefore provides more accurate estimation of PCF migration rates.

\begin{table}[pht]
\caption{Migration rates of PCFs.
\label{migrate}}
\centering
\begin{tabular}{lcccr}
\\
\tableline\tableline
               & Cycle 21 & Cycle 22 & Cycle 23 & Cycle 24 \\
               & deg/rotation & deg/rotation & deg/rotation & deg/rotation \\
\tableline
North Poleward & 0.65 $\pm$ 0.07    & 0.40 $\pm$ 0.06    & 0.67 $\pm$ 0.10    & 0.55 $\pm$ 0.10    \\
South Poleward & 0.43 $\pm$ 0.05    & 0.41 $\pm$ 0.05    & 0.42 $\pm$ 0.05    & 0.63 $\pm$ 0.08    \\
N-S average    & 0.54     & 0.41     & 0.55     & 0.59     \\
N-S difference & 0.22     & -0.01     & 0.25     & -0.08    \\
\tableline
\end{tabular}
%\end{center}
\end{table}
\vspace{-1.2em}
{Notes: The migration rates are derived by fitting the PCF positions as a function of time.}\\

\begin{figure}
\centering
\includegraphics[scale=0.80]{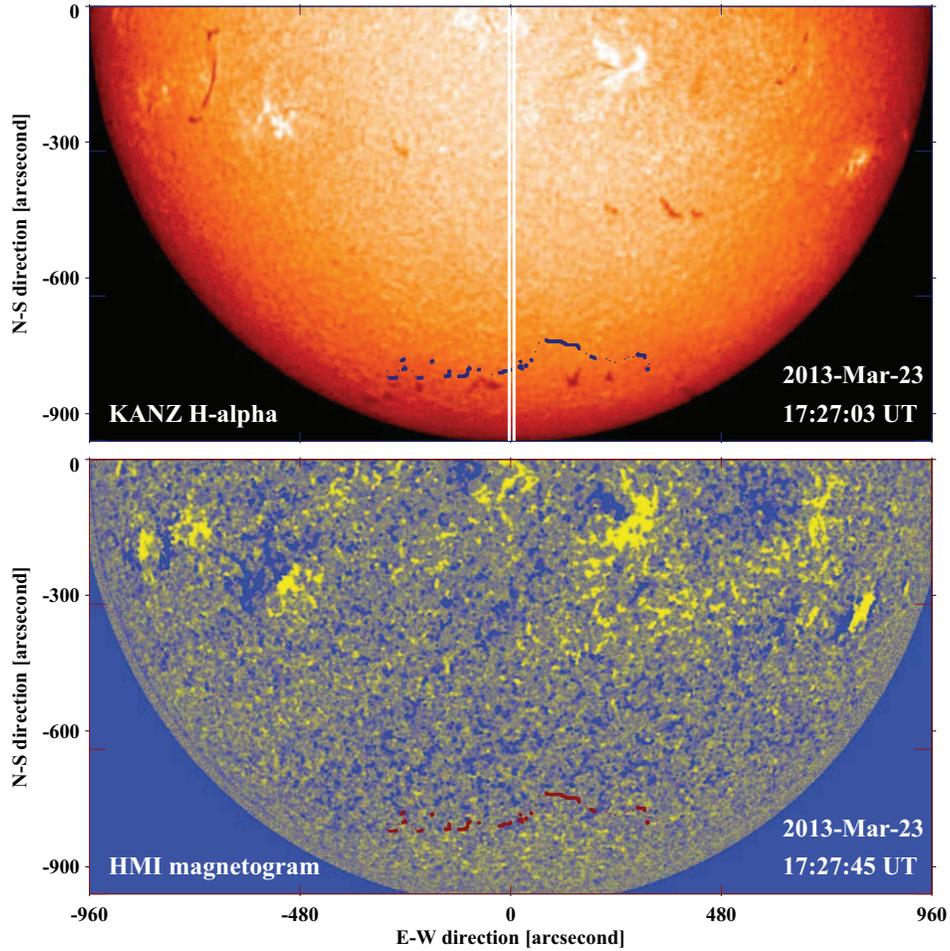}
\caption{Top panel: KSO image showing a PCF near south pole (dark feature near the bottom), taken on 2013-March-23. The blue dashed curve indicates the magnetic PIL. Two white lines outline the central meridian of the Sun, helping to measure the central position of the PCF. Bottom panel: SDO/HMI LOS magnetogram. The same PIL are shown in red color. Here, the blue color indicates magnetic fields in negative polarity and yellow color represents positive polarity.
\label{measure}}

\end{figure}

For each PCF, we measure the coordinates of its left, right ends and
center, in which the center coordinates is used to represent the position of the PCFs.
Figure~\ref{measure} shows a PCF near the north pole (lower panel) and the corresponding SOHO/MDI magnetogram (upper panel), demonstrating that the PCF is co-spatial with the polar PIL (dotted blue curve in lower panel). Two white vertical lines are drawn on the H$\alpha$ map, right next to the central meridian of the solar disk, to help locating the filament position relative to the central meridian. For segmented PCFs, interpolation is done visually to find the intersection of a PCF and the central meridian. The measurement error is limited to 5\arcsec\ in E-W orientation and dose not exceed half width of the filament, which is usually about 5\arcsec\ to 20\arcsec. The pixel coordinates are converted into heliographic longitudes and latitudes. Figure~\ref{mag} shows the PCF latitudes as a function of time (in the unit of Carrington rotation number), using green circular bullets. First of all, clear periodic migrations of PCFs are seen during all of the 4 cycles, which is consistent with previous reports, such as in \citet{Cliver2014} and \citet{Gopalswamy2018}. On the other hand, the migration pattern is more outstanding using the manual detection method than that using automatic detection method \citep{Hao2015, Chatterjee2017} in the overlapping periods. The slopes of red trend lines indicate the migration rates, which are listed in Table~\ref{migrate}. Furthermore, we compare the PCFs migration with the polar magnetic field strength (blue and red dots) and a good correlation is seen in Figure~\ref{mag}. Table~\ref{appearance} lists the number of rotations without any PCFs found and its percentage compared to the total number of rotations in each cycle. The inferred key results are summarized as following:

\begin{enumerate}

\item[1.] On average, the migration rates are about 0.4 to 0.7 degree per rotation, which is similar to the results in \citet{Altrock2014}.

\item[2.] The N-S asymmetry of migration speed is clear in cycles 21, 23 and 24, and not clear in cycle 22. In particular, the northward migration speed was faster than southward speed in cycle 23. However, in cycle 24, this relationship reversed. Such N-S asymmetry in cycles 23 and 24 agree with the results in \citet{Gopalswamy2016}.

\item[3.] There are more PCFs seen in cycles 21 and 22 than that in cycles 23 and 24. The N-S asymmetry of PCF occurrence is not as significant as PCFs migration rate.

\item[4.] The PCFs start poleward migration right after the new cycle begins and reach their maximum latitude near the solar maximum when the pole field strength approaches to zero.

\end{enumerate}

\begin{table}[pht]
\caption{Number of rotations without PCFs. \label{appearance}}
\centering
\begin{tabular}{lcccr}
\\
\tableline\tableline
               & Cycle 21 & Cycle 22 & Cycle 23 & Cycle 24 \\
\tableline
North & 75   & 53 & 103 & 84 \\
South & 69   & 62 & 107 & 76 \\
Total Rotations & 141   &  133   &  165  &  108 \\
N\%   & 53\%    & 40\%     & 62\%     & 78\%     \\
S\%   & 49\%    & 47\%     & 65\%     & 70\%     \\
N-S\% & 4\%     & -7\%     & -3\%     & 8\%      \\
\tableline
\end{tabular}
%\end{center}
\end{table}
\vspace{-1.2em}
{Notes: The first two rows list the number of Carrington rotations where no PCF
was found. The third row shows the total number of rotations (duration) in each solar cycle. The fourth and fifth rows show the percentages of rotations without PCF over the duration of each cycle (row 1 and 2 divide by row 3).}\\

\section{Discussion}

In this study, we identified about 700 PCFs from H$\alpha$ images obtained from 1973 to 2018, covering solar cycles 21 to 24, based on visual detections. Clear poleward migration patterns of PCFs are seen and the estimated migration rate is about 0.4 to 0.7 degree per Carrington rotation, corresponding to $\sim$27 days. We found good correlation between the periodic variation of PCFs and the polar magnetic fields, obtained from NSO, SOHO/MDI and SDO/HMI. Usually the length of filaments are estimated as an characteristics \citep{Hao2015}. However, it is meaningless to measure the length of PCFs, as many of them are circulating the poles or have highly segmented shape. Therefore, the size (length) of PCFs is not taken into account in this study.

\begin{figure}
\centering
\includegraphics[scale=1]{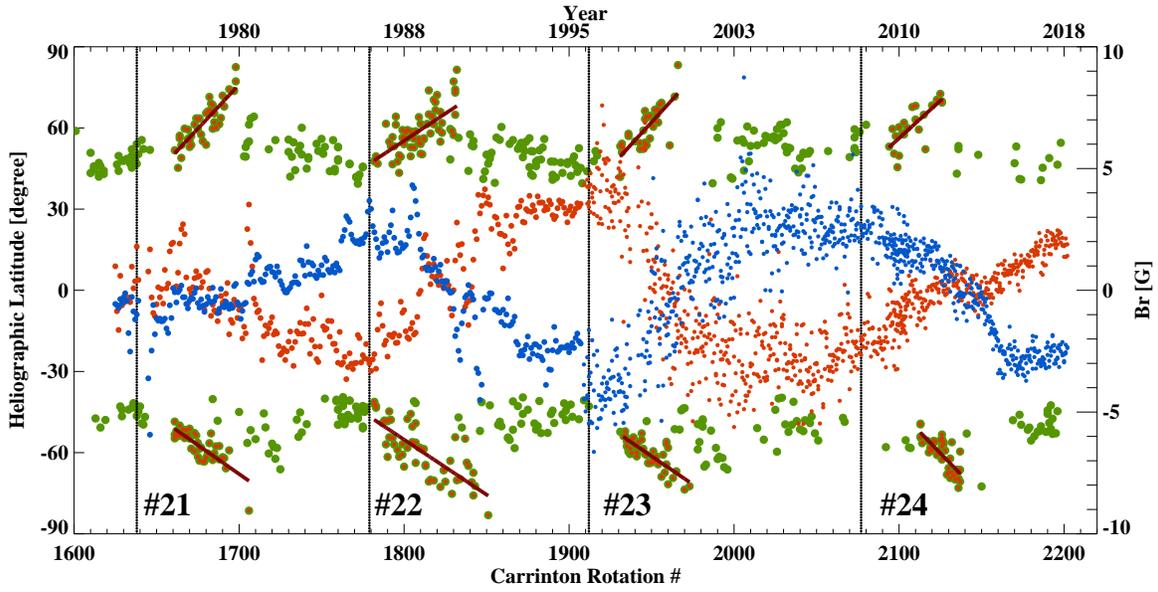}
\caption{Comparison of PCF (green dots) migration and temporal variation of polar magnetic fields (blue and red dots for northern and southern polar fields, respectively). Trend lines of PCF migrations are plotted as red lines, during the ascending phases (green dots with red centers). Black vertical lines indicate the beginning of the solar cycles, according to Sunspot Index and Long-term Solar Observations (SILSO) at Royal Observatory of Belgium.
\label{mag}}

\end{figure}

We found that cycles 23 and 24 are different from the preceding cycles 21 and 22, in terms of the variation of polar field and appearance of PCFs. An examination of Figure~\ref{mag} suggests that there are fewer PCFs in cycles 23 and 24. PCFs usually last much longer than AR filaments, they may be double counted during consecutive rotations. Therefore, counting the numbers of PCFs directly does not have much physical meaning. To circumvent this problem, we counted the number of rotations without PCF is seen on the disk and the results is shown in Table~\ref{appearance}. It is obvious that there were less PCFs in cycles 23 and 24, in which the absolute number of non-PCF rotations reaches maximum in cycle 23 and the relative percentage reaches maximum in cycle 24. In addition, the polar magnetic fields vary in an `abnormal' way in cycle 24. The fields increased rapidly after the solar maximum (polar fields reduce to near zero) and turned into a constant phase (horizontal, see Figure~\ref{mag}) in the south pole. \citet{Gopalswamy2016} found that in cycle 24, the polarity reversal was completed earlier in the south pole than that in the north pole, which was in an opposite order comparing to cycle 23. Our results show that the migration speed of the north PCFs is faster than the speed of south PCFs in cycles 21 and 23 (cycle 22 is not clear). But this relation changed in cycle 24, in which the south PCFs migrated faster than the north PCFs. \citet{Gopalswamy2016} found that the south pole reversed faster, indicating the PILs on the boundary of south pole shrank faster in cycle 24. As a consequence, PCFs should migrate faster as found in our results.

The above results of PCFs in the past four cycles suggest that PCFs can provide additional clues in understanding the variation of magnetic fields, especially the polar fields, among different solar cycles. Extended studies, including more cycles, are motivated to advance our understanding of relationship between PCFs and polar fields and their variations along with solar cycles.

Obtaining the excellent data would not have been possible without the help of the KSO and BBSO teams. BBSO operation is supported by NJIT. We would like to thank the anonymous referee for the very valuable comments in improving this work. This work is supported by NSF grant AGS 1620875.

\bibliographystyle{apj}
%\bibliography{E:/yxu/reference/reference}

\end{document}